# Chemical potential dependent gap-opening at the Dirac surface states of $Bi_2Se_3$ induced by aggregated substitutional Cr atoms

Cui-Zu Chang, [1, 2, 3] Peizhe Tang, [1,2] Yi-Lin Wang, [3] Xiao Feng, [1, 2, 3] Kang Li, [2, 3]
Zuocheng Zhang, [1,2] Yayu Wang, [1,2] Li-Li Wang, [1, 2, 3] Xi Chen, [1, 2] Chaoxing Liu, [4]
Wenhui Duan, [1, 2†] Ke He, [1, 2, 3‡] Xu-Cun Ma, [1, 2, 3] and Qi-Kun Xue[1,2]

[1] *State Key Laboratory of Low-Dimensional Quantum Physics, Department of Physics, Tsinghua University, Beijing 100084, China*

[2] *Collaborative Innovation Center of Quantum Matter, Beijing, China*

[3] *Beijing National Laboratory for Condensed Matter Physics, Institute of Physics, Chinese Academy of Sciences, Beijing 100190, China*

[4] *Department of Physics, The Pennsylvania State University, University Park, Pennsylvania 16802-6300, USA*

**With angle-resolved photoemission spectroscopy, gap-opening is resolved at up to room temperature in the Dirac surface states of molecular beam epitaxy grown Cr-doped $Bi_2Se_3$ topological insulator films, which however show no long-range ferromagnetic order down to 1.5 K. The gap size is found decreasing with increasing electron doping level. Scanning tunneling microscopy and first principles calculations demonstrate that substitutional Cr atoms aggregate into superparamagnetic multimers in $Bi_2Se_3$ matrix, which contribute to the observed chemical potential dependent gap-opening in the Dirac surface states without long-range ferromagnetic order.**

Time-reversal (TR) invariant topological insulators (TIs) are a new class of topological matters whose topologically non-trivial property is induced by spin-orbit coupling and protected by TR symmetry [1]. A three-dimensional (3D) TI is characterized by gapless surface states with Dirac-cone-shaped band dispersion around TR invariant points of surface Brillouin zone [2]. Ferromagnetic ordering can break the TR symmetry of a 3D TI and open a gap at the Dirac point of the surface states [3]. Many novel quantum phenomena predicted in 3D TIs, e.g. topological magneto-electric effect, image magnetic monopoles, and quantum anomalous Hall effect [3-6], result from the magnetically gapped Dirac surface states (DSS).

Doping magnetic impurities is a convenient approach to induce ferromagnetic order in a semiconductor/insulator if exchange coupling between impurities far from each other could be built [7,8]. Several long-distance coupling mechanisms supporting long-range ferromagnetic order in magnetically doped TIs have been proposed [8-10]. Many experimental studies, from angle-resolved photoemission spectroscopy (ARPES) to transport measurements, have been done on magnetically doped $Bi_2Se_3$ family TIs: $Bi_2Se_3$, $Bi_2Te_3$, and $Sb_2Te_3$, which are by far the most popular TI materials [11-16]. The experimental results, however, are inconsistent between different materials and different measurement methods. In magnetically doped $Bi_2Te_3$ and $Sb_2Te_3$, transport and magnetization measurements have shown clear and consistent long-range ferromagnetic order [12,13] which leads to the observation of the quantum anomalous Hall effect, a representative quantum phenomenon predicted in magnetic TIs [6]. Nevertheless, magnetically induced gap-opening at DSS could not be resolved by ARPES in these materials due to the low Curie temperature ($T_C$) and small gap size. $Bi_2Se_3$ is the most-studied and presumably the best material of $Bi_2Se_3$ family TIs for its largest bulk gap and the Dirac point residing right in the gap [11]. Magnetic-doping-induced gap-opening at DSS has been reported in several ARPES studies on various magnetically doped $Bi_2Se_3$, with the observed gap size up to one hundred meV, implying a strong ferromagnetism [14,15]. However transport and magnetization measurements could only show very small or even no magnetic hysteresis in the systems, making the existence of long-range ferromagnetic order questionable and preventing further progress to various quantum effects [14-16]. In this work, aiming on solving the puzzle in magnetically doped TIs, we have investigated the relation between the magnetism and surface state gap of molecular beam epitaxy (MBE)-grown Cr-doped $Bi_2Se_3$ by combining ARPES, transport measurements, scanning tunneling microscopy (STM), and first-principles calculations. Our results reveal that substitutional Cr impurities in $Bi_2Se_3$ form superparamagnetic multimers, each with rather high $T_C$ each but without long-range ferromagnetic order in whole. It leads to gapped DSS shown in ARPES in the absence of macroscopic ferromagnetism in transport and magnetization measurements.

Cr-doped $Bi_2Se_3$ films are prepared by MBE on graphene-terminated silicon carbide (0001) (for ARPES and STM measurements) or sapphire (0001) substrates (for transport measurements) [17]. *In situ* ARPES measurements are taken with a Scienta SES-2002 analyzer and a Gammadata helium discharging lamp producing unpolarized He-Iα photons ($h\upsilon$ = 21.218 eV). STM experiments are performed at 4.2 K with a Unisoku low-temperature STM system. The density functional theory calculations are carried out using the Vienna *ab initio* simulation package (VASP) with Perdew-Burke-Ernzerhof type generalized gradient approximation for exchange and correlation [17-19].

We basically use 8 quintuple (QL) thick films in this study. Such a thickness is more than enough for a $Bi_2Se_3$ film to enter the 3D TI regime, since the gap induced by hybridization between the top and bottom DSS can only be observed in 5 QL and thinner films [20]. On the other hand, compared with thicker films, the DSS peaks of a 8 QL $Bi_2Se_3$ film are more clearly resolved in ARPES thanks to the more uniform film thickness and quantization of bulk bands at low thickness [20]. However the phenomena observed from 8 QL thick films shown below can also be repeated in thicker samples [17].

Figure 1 displays the RT-measured ARPES grey-scale bandmaps (upper panels) and the corresponding energy distribution curves (EDCs) (lower panels) of a pure $Bi_2Se_3$ film and a Cr-doped $Bi_2Se_3$ ($Bi_{1.96}Cr_{0.04}Se_3$) film, both of which are 8 QL thick. In the spectra for the pure $Bi_2Se_3$ film [Figs. 1(a) and 1(c)], the DSS and parabolic quantum-well states are clearly observed. The Dirac point manifests itself as a distinct protrusion in the spectrum around $\bar{\Gamma}$ point [indicated by the red arrow in Fig. 1(c)], demonstrating gapless nature of the surface states. With a small amount of Cr impurities doped in $Bi_2Se_3$ (2% of the cations) [Figs. 1(b) and 1(d)], the spectrum near the Dirac point changes from a protrusion into a dip [indicated by the red arrow in Fig. 1(d)], clearly showing a gap opened here with the gap size about 75 meV. The Cr-doping induced gap-opening can also be observed in thicker films and higher Cr concentrations, with gap size increases with increasing Cr content [17].

The 75 meV gap observed at RT suggests that Cr impurities form a rather strong ferromagnetism with $T_C$ above RT. But, it is not supported by transport measurements. Figures 2(a) and 2(b) display magnetic field ($\mu_0 H$) dependence of the Hall resistance ($R_{yx}$) and longitudinal conductivity of an 8 QL $Bi_{1.96}Cr_{0.04}Se_3$ film, respectively at varied temperature. Above 10 K, $R_{yx}$ exhibits a linear relationship with $\mu_0 H$, and the magneto-conductance experiences a weak anti-localization behavior, both are typical properties of a non-magnetic TI [Fig. 2(b)] [21]. Below 10 K, a non-linear Hall effect gradually develops, accompanied by the evolution of magneto-conductance to weak localization behavior, implying formation of some magnetism [22]. However, long-range ferromagnetic order is absent since no hysteresis is observed in any case.

An enough high Cr doping level can induce a phase transition of $Bi_{2-x}Cr_xSe_3$ from a TI to a topologically trivial insulator, which will destroy the DSS and the ferromagnetism held by van Vleck mechanism [23]. The Cr concentration of the samples studied here however is far below the critical value for such a topological phase transition ($x \sim 0.15$). Furthermore the surface bands in the present samples are only gapped around Dirac point, instead of being completely removed, which also excludes the topologically phase transition.

To probe the origin of the magnetic-doping-induced gap-opening of DSS, we have systematically studied the chemical potential ($\mu$) dependence of gap size ($\Delta$), which reflects the ferromagnetic exchange energy [1], for different magnetic coupling mechanisms have different $\mu$ dependences [8-10]. The $\mu$ of $Bi_2Se_3$ can be tuned by doping Mg impurities [14,24] and by taking ARPES measurements at different temperatures through surface photovoltage effect [20,25]. Figures 3(a)-3(c) show the ARPES EDCs of 8 QL $Bi_{1.96}Mg_yCr_{0.04}Se_3$ films with different Mg concentrations ($y$) measured at 150 K. In the film without any Mg doping [Fig. 3(a)], the mid-gap energy ($E_0$) is located at −303 meV, and $\Delta$ is about 59 meV. The value of $\Delta$ is estimated by fitting the spectrum near the gap at $\bar{\Gamma}$ point with double Lorentzian peaks, as shown in the inset of Fig. 3(g). Doping Mg to $y$ = 0.0015 shifts $E_0$ to −275 meV, and $\Delta$ is increased to 85 meV [Fig. 3(b)]. As $y$ is increased to 0.003 [Fig. 3(c)], $E_0$ is further shifted to −208 meV, whereas $\Delta$ is enhanced to 146 meV. Figures 3(d)-3(f) show the ARPES EDCs of an 8 QL $Bi_{1.96}Cr_{0.04}Se_3$ film measured at different temperatures. At 80 K, the film is heavily electron-doped with $E_0 \sim$ −403 meV due to photo-induced electron-doping [20,25]. The gap becomes so small that the surface states almost return to a gapless Dirac cone. At 150 K, with alleviative photo-doping, $E_0$ is shifted to −303 meV, whereas $\Delta$ is increased to 59 meV. At RT, $E_0$ reaches $\sim$ −280 meV, whereas $\Delta$ becomes $\sim$75 meV, which can be clearly resolved. This $\mu$ dependence of $\Delta$ is similar to that shown in varying Mg doping experiments. We have performed many ARPES measurements on samples with the same Cr doping level but different Mg concentrations and at different temperatures. The data of $\Delta$ and $|E_0|$ obtained from the measurements (not limited to the spectra shown above) are plotted in Fig. 3(g). It is clear that $\Delta$ increases monotonously with decreasing $|E_0|$, regardless of the methods used to change $\mu$.

At the first sight of the above data, one is tempted to attribute the gap-opening to the Ruderman-Kittel-Kasuya-Yosida (RKKY) - like magnetic coupling mediated by the DSS [9]. This surface RKKY ferromagnetism becomes stronger as Fermi level approaches Dirac point, and may not be detected by bulk-sensitive transport measurements since it only exists near surface region. However it is hard to believe that the surface RKKY ferromagnetism could show $T_C$ above RT [9]. Furthermore, we have tried surface-sensitive magnetic optic Kerr effect measurements on the Cr-doped $Bi_2Se_3$ films down to 10 K, which also fails to show any ferromagnetic signal. So it is unlikely that the surface state gap is induced by surface RKKY ferromagnetism.

We have performed STM measurements on Cr-doped $Bi_2Se_3$ to identify the configuration of Cr impurities in $Bi_2Se_3$ matrix. Figure 4(a) shows a typical STM image of an 8 QL $Bi_{1.96}Cr_{0.04}Se_3$ film. The terrace is scattered with defects (dark regions) resulting from Cr doping. Figure 4(b) shows a high-resolution STM image around one of the defects, which exhibits a polygonal shape in register with the surrounding lattice of Se atoms. The minimum-sized defects observed have an equilateral triangle shape covering three Se atoms, as shown in Fig. 4(c) (indicated by a black centered triangle). Similar to the look of magnetic impurities observed in Mn-doped $Bi_2Te_3$ and Cr-doped $Sb_2Te_3$ [12,13], the triangle-shaped defect can be attributed to a single substitutional Cr atom occupying sub-surface Bi site (the center of the triangle), reducing the local density of states (DOS) of the three above Se atoms (the corners of the triangle). Larger defects can thus be attributed to multimers of several substitutional Cr atoms aggregated together. For example, the defect shown in Fig. 4(b) is composed of 8 substitutional Cr atoms, as indicated by the 8 black centered triangles. It is different from Mn-doped $Bi_2Te_3$ and Cr-doped $Sb_2Te_3$ in which most defects shown in STM are triangle-shaped single substitutional ones away from each other [12,13].

We have used the first-principles calculations to investigate the properties of the multimers of substitutional Cr atoms. The simulated multimers contains three substitutional Cr atoms (trimmers). There are two possible configurations of Cr trimmers which are hereafter referred as $C_1$ and $C_2$ respectively, depending on the Se atoms locating in their centers [see the structural models shown in Figs. 4(d), 4(e), 4(h), and 4(i)]. In a $Bi_2Se_3$ QL, Bi and Se atomic layers stack in the sequence of Se1-Bi-Se2-Bi-Se1 along the [111] direction [11]. $C_1$ and $C_2$ indicate the Cr trimmers centered on Se atoms from Se2 [$Se^I$ atom in Figs. 4(d) and 4(e)] and Se1 layers [$Se^{III}$ atom in Figs. 4(h) and 4(i)], respectively. The two types of trimmers have distinct charge density distributions, leading to different STM images [as shown in Figs. 4(f) and 4(j)], both of which have been observed [see Figs. 4(g) and 4(k)].

The calculated formation energies of $C_1$ and $C_2$ trimmers under Se-rich condition (the usual condition in MBE growth of $Bi_2Se_3$) are 1.155 eV and 1.158 eV, respectively, both are much lower than that of isolated Cr substitution (1.323 eV) [17]. It explains why substitutional Cr atoms favor aggregation over dispersed distribution in $Bi_2Se_3$. The calculated projected DOS of $Se^I$ $4p$, $Se^{II}$ $4p$, and Cr $3d$ orbitals of $C_1$ are shown in Figs. 4(l)-4(n), respectively. Due to crystal-field splitting, Cr $3d$ orbitals are split into $t_{2g}$ and $e_g$ states: three electrons occupy the localized $t_{2g}$ states and the unoccupied $e_g$ state strongly hybrids with the $p_y$ orbitals of $Se^I$ and $Se^{II}$ atoms. The strong $p$-$d$ hybridization can induce ferromagnetism in each trimmer, as observed in $Cd_{1-x}Mn_xTe_3$ [26] and $Zn_{1-x}Cr_xTe_3$ [27], through ferromagnetic superexchange mechanism [28]. $C_2$ has the similar situation with $C_1$ [17]. Energy-mapping analysis [29] shows that the ferromagnetic coupling strengths between two neighboring Cr ions are 24 meV and 29 meV for $C_1$ and $C_2$, respectively, corresponding to $T_C$ near RT [30]. In larger multimers, higher $T_C$ is expected. The calculated magnetic moment is 3 $\mu_B$ per Cr ion, consistent with SQUID measurements. The calculated magnetic anisotropy is perpendicular to film plane with the energy $K \sim 7.06 \times 10^6$ ergs/cm$^3$.

Non-uniform distribution of magnetic impurities is detrimental to long-range ferromagnetic order. For example, surface RKKY ferromagnetic coupling requires that the inter-impurity distance be at least near the Fermi wavelength. Longer distance leads to an exponential decrease of $T_C$ [9]. From our ARPES data, the Fermi wavelength of DSS of $Bi_{1.96}Cr_{0.04}Se_3$ film is around 1 nm, much shorter than the average inter-multimer distance (~ 6 nm) observed by STM. Thus ferromagnetic coupling between Cr multimers is negligible here, which means a superparamagnetic system. The Hall traces indeed exhibit a typical superparamagnetic behavior with $T_B \sim$ 10 K [Fig. 2(a)]. From the $T_B$ and the calculated magnetic anisotropy energy ($K \sim 7.06 \times 10^6$ ergs/cm$^3$), the average size $d$ of multimers is estimated to be ~ 1 nm [27,31], which is in good agreement with the STM observations.

Although in superparamagnetism, the magnetic moments of Cr multimers keep flipping between two spin states due to thermal fluctuation, the strong perpendicular magnetic anisotropy guarantees the magnetization perpendicular to the surface most of the time, no matter in spin-up or spin-down states. The DSS can be gapped near multimers in both of the spin states. In ARPES, the measured photoelectron intensity is the sum of the time-averaged signals from both gapped and gapless regions, and thus can exhibit gap-induced suppression (see the simulated result in Ref. [17]).

Charging electrons into the Cr multimers can suppress their ferromagnetism according to our calculations [17]. For $C_1$, the charged electrons mainly locate at Se$^I$ and Cr atoms and, due to Coulomb repulsion, induce a significant structural distortion: the bond-lengths of Cr-Se$^I$ and Cr-Se$^{II}$ increase from 2.67 Å and 2.61 Å to 2.93 Å and 2.63 Å, respectively, and the bond angles of Cr-Se$^I$-Cr and Cr-Se$^{II}$-Cr change from 94.6 ° and 97.6 ° to 86.8 ° and 99.7 °, respectively. This distortion weakens the ferromagnetic coupling between neighboring Cr ions, from 24 meV to 16 meV, similar to the Zn$_{1-x}$Cr$_x$Te$_3$ case [27]. For $C_2$, the distribution of charged electrons is less localized and the induced Coulomb repulsion is not so strong; thus the ferromagnetism shows weaker $\mu$ dependence. However Cr multimers larger than trimmers usually contain at least one $C_1$ [17], which can show $\mu$ dependence of their ferromagnetism. Thus the observed $\mu$ dependence of $\Delta$ can be attributed to the variation in ferromagnetism.

It has long puzzled the researchers of TIs that some magnetically doped TIs, especially that of $Bi_2Se_3$, can exhibit rather large magnetic doping induced gap at their DSS but cannot show good long-range ferromagnetic order in transport and magnetization measurements, let alone the predicted topological quantum effects [14-16]. The present study demonstrates that the seemingly inconsistent observations come from short-range ferromagnetic order of aggregated magnetic impurities. Non-uniform distribution of magnetic impurities is a common and crucial issue in traditional magnetically doped semiconductors/insulators, but has not yet received as much attention in magnetically doped TIs. Our results first show how the DSS and magnetism of a TI are influenced by non-uniformly distributed magnetic impurities, which is instructive not only for understanding of the existing magnetically doped TI materials but also for the search for new ones. Furthermore, the magnetic nanodots embedded in a TI revealed here represents a new class of low-dimensional magnetic system in which the electronic properties of host material can be tuned by chemical potential via magnetism of nanodots. It promises new concept field effect devices for future electronic/spintronic applications.

C. C. and P. T. contributed equally to this work. We thank Z. Fang, X. Dai, K. Chang, X. L. Qi, S. C. Zhang, and L. Fu for helpful discussions. This work was supported by the National Natural Science Foundation of China, the Ministry of Science and Technology of China, and the Chinese Academy of Sciences.

*dwh@phys.tsinghua.edu.cn

†kehe@tsinghua.edu.cn

**Figure Captions:**

**FIG. 1** (color online). ARPES grey-scale bandmaps (along the $\bar{\Gamma}$-$\bar{K}$ direction) (a,b) and the corresponding EDCs (c,d) of an 8 QL $Bi_2Se_3$ film (a,c) and an 8 QL $Bi_{1.96}Cr_{0.04}Se_3$ film (b,d) taken at RT.

**FIG. 2** (color online). Hall traces (a) and magneto-conductivity (b) of an 8 QL $Bi_{1.96}Cr_{0.04}Se_3$ film measured at varied temperatures.

**FIG. 3** (color online). (a-c) The ARPES EDCs of 8 QL $Bi_{1.96}Mg_yCr_{0.04}Se_3$ films taken at ~150 K with $y = 0$ (a), $y = 0.0015$ (b), and $y = 0.003$ (c). (d-f) The ARPES EDCs of an 8 QL $Bi_{1.96}Cr_{0.04}Se_3$ film taken at ~80K (d), ~150K (e), and ~300K (f). (g) The relation between $\Delta$ and $|E_0|$ obtained from different ARPES measurements. The inset shows the fitting result of the $\bar{\Gamma}$ point spectrum of an 8 QL $Bi_{1.96}Cr_{0.04}Se_3$ film with double Lorentzian function.

**FIG. 4** (color online). (a-c) STM images of 50 nm × 50 nm (a), 3.8 nm × 3.8 nm (b), and 4 nm × 4 nm (c) taken from an 8 QL $Bi_{1.96}Cr_{0.04}Se_3$ film. (d,e,h,i) Top (d,h) and side (e,i) views of the geometrical structures of $C_1$ (d,e) and $C_2$ (h,i) in a $Bi_2Se_3$ QL, respectively. (f,g,j,k) Simulated (f,j) and measured (g,k) STM images of $C_1$ (f,g) and $C_2$ (j,k), respectively. (l-n) PDOS of $Se^I$, $Se^{II}$ and Cr atoms in $C_1$.

**FIG. 1**

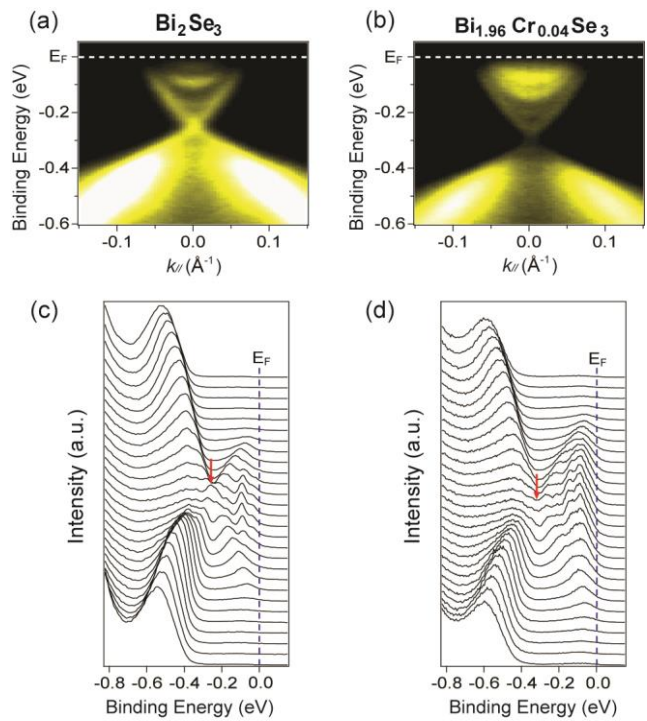

F

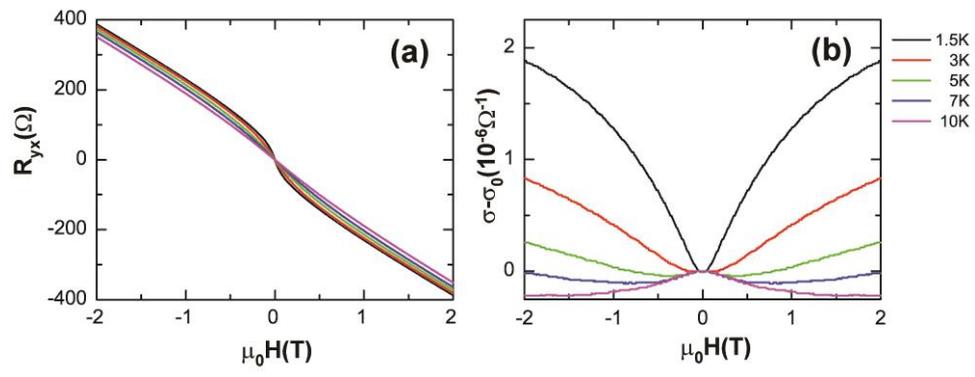

**FIG. 3**

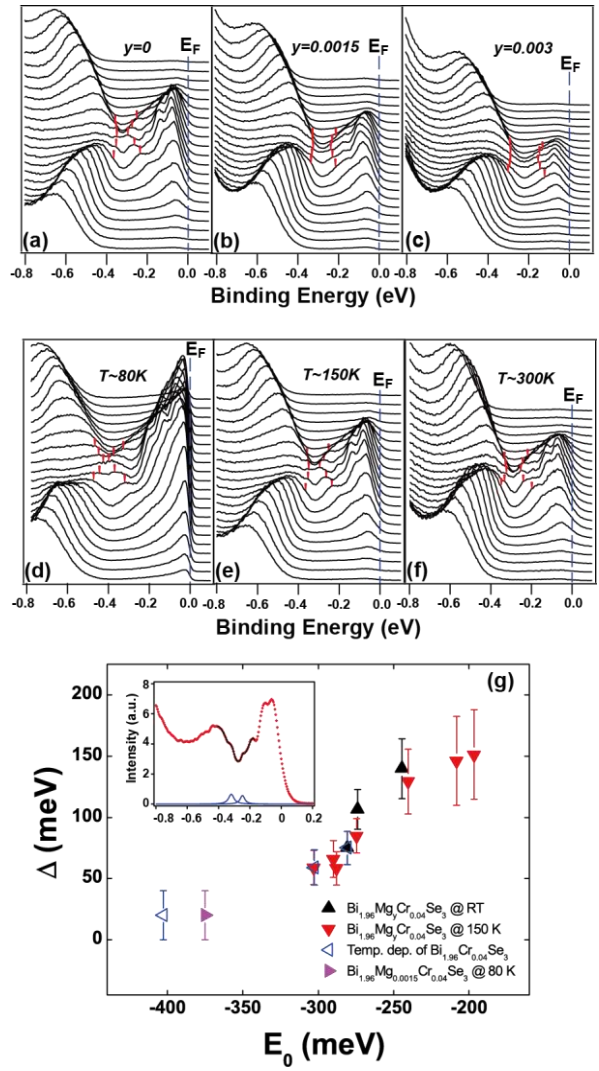

# FIG. 4

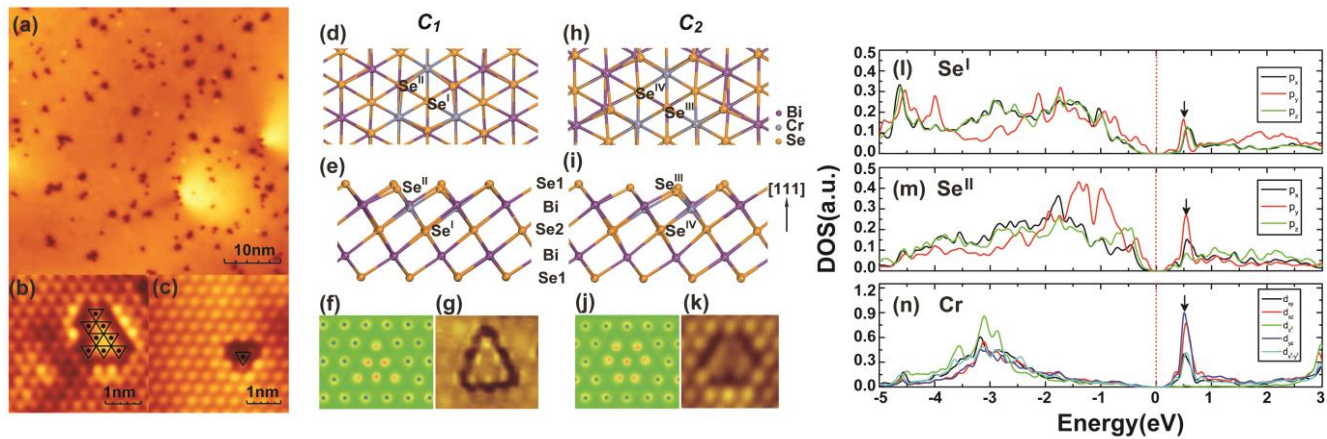